\documentclass{ametsocV6.1_noln}
\usepackage{amsmath}
\usepackage{amssymb}
\usepackage{graphicx}
\usepackage{multirow}

\DeclareMathOperator*{\argmax}{arg\,max}

\title{Investigating How the Fractions Skill Score and Brier Divergence Skill Score Reflect Forecast Error}

\authors{Bobby Antonio\aff{a}\correspondingauthor{bobby.antonio@physics.ox.ac.uk}
}
\affiliation{\aff{a}{Department of Physics, University of Oxford, UK}
}

\abstract{
Meaningful scores for forecast verification are essential for developing reliable forecasts, and there has been much effort to develop scores that align well with human perceptions of forecast quality. Whilst many of these scores have intuitive interpretations, relatively little is known about how these scores rank different forecasts, and how scores reflect forecast error. We theoretically explore the behaviour of two scores that fall within the `neighbourhood' paradigm of spatial verification; the Fractions Skill Score (FSS) and Brier Divergence Skill Score (BDnSS). We investigate how each score ranks forecasts with two types of error; errors in the mean frequency (corresponding to intensity or shape errors) and errors in the standard deviation (corresponding to errors in spatial structure, such as blurring or excess noise). We find that under many situations the FSS assigns higher scores to forecasts that over-predict mean frequency, thus theoretically confirming the need to use the FSS with percentile thresholds. Both scores generally assign higher scores to forecasts with lower neighbourhood standard deviation, a reflection of the `double penalty' problem; however, we observe that size of this effect is larger for the BDnSS than the FSS, showing that the FSS under some situations is less susceptible to the double penalty problem than the BDnSS.
}

\begin{document}

\maketitle

\statement
Scores that summarise weather forecast performance are valuable tools in developing accurate numerical weather models. Many scores have been developed, targeting different aspects of forecast performance, but there is still much we can learn about how these scores behave, and what insights they provide to a forecaster. We investigate two scores within the `neighbourhood' paradigm of forecast verification, and reveal novel theoretical properties of these scores under particular types of error. This analysis provides more clarity on how best to use these scores for forecast model improvement and understanding.

%
\section{Introduction}

Forecast verification is a crucial yet challenging aspect of numerical weather model development. Simple and commonly used metrics like mean square error are known to suffer from the `double penalty' problem \citep{wilks_forecast_2019}, whereby distortions such as blurring that reduce forecast realism can artificially drive score improvements. A range of verification methods have therefore been proposed to provide more nuanced summaries of forecast performance (e.g. \cite{casati_forecast_2008, gilleland_intercomparison_2009,ebert_progress_2013}). The high number of verification methods available makes it challenging to choose the most appropriate score for a given situation, particularly when relatively little is known about their properties, and the assumptions underpinning them.

Therefore, it is important that we perform `metaverification', in order that a forecaster can understand limitations of these scores and select the most appropriate score for the task. This is also important as machine learning models become more prevalent in weather prediction, so that we may identify scores which are most suitable as loss functions \citep{lagerquist_can_2022,ebert-uphoff_cira_2021}, and where simple loss functions such as mean-squared error encourage overly smooth forecasts \citep{subich_fixing_2025}.

One approach to metaverification is to have human forecasters assess the relative ranking that different scores assign to specially chosen case studies. This can provide a high quality of assessment, however it is limited in scale and can be sensitive to the subjectivity of the forecasters. A second approach is to synthetically apply certain perturbations (such as displacement, deformation, blurring, or multiplicative bias) to forecast fields or simplified objects, and compare the verification scores to the magnitude of the perturbations. This provides a more systematic way of comparing verification scores, but may not truly reflect how these scores behave with more realistic forecasts and observations. These two methods were comprehensively applied in the spatial verification intercomparison project (ICP, \cite{gilleland_intercomparison_2009}) and in other works such as \cite{skok_spatial_2025}.

A third approach to metaverification is to look for theoretical properties of the verification methods. This can be challenging, as the observations and forecasts create complex spatial patterns that are hard to describe theoretically, and the verification methods themselves may not be easy to investigate. Examples of this approach are the analytical expressions of the Fractions Skill Score for idealised forecasts \citep{skok_analysis_2016,skok_analysis_2015}, and limiting cases over small domains \cite{mittermaier_meta_2021}, or the effects of forecast error on contingency tables \citep{manzato_behaviour_2017}. It is clearly worthwhile pursuing this approach if possible, in order to obtain some general properties of different scores. 

In this work, we pursue this third approach, and provide new insights into the theoretical properties of two scores within the `neighbourhood' paradigm of spatial verification \citep{ebert_fuzzy_2008}. For neighbourhood verification methods, forecasts and observations are aggregated over a neighbourhood, with extensions that include aggregation over neighbouring time steps or ensemble members. We can equivalently think of this neighbourhood aggregation as spatial smoothing, or creating a pseudo-ensemble by resampling in space and/or time \citep{theis_probabilistic_2005}. The intuition behind this technique is that it will reduce the effects of small displacement errors in the forecasts, thereby mitigating the double penalty problem. However, it is not clear whether all neighbourhood scores are equally effective at dealing with the double penalty problem or not, which motivates this study. 

We investigate two neighbourhood scores; the Fractions Skill Score (FSS, \cite{roberts_assessing_2008, roberts_scale-selective_2008}), which is perhaps the most widely used of the neighbourhood scores, and the recently introduced Brier Divergence Skill Score (BDnSS, \cite{stein_evaluation_2024}). To analyse these scores, we present a simple framework to explore how they are affected by different errors. First, the scores are expressed in terms of summary statistics, using a similar decomposition to \cite{antonio_derive_2025}. Then the scores are expressed in terms of multiplicative errors in the mean and standard deviation of the neighbourhood fractions. From these forms, we vary only one of the multiplicative errors with all other terms kept fixed, and look for the multiplicative error value which maximises the score. This assesses whether or not the score can be improved by deviating the properties of the forecast away from the observations. This follows a similar philosophy to the idea of `hedging', introduced in the context of probabilistic forecasts \citep{murphy1978hedging}, whereby a score is said to be susceptible to hedging if it encourages a forecaster to submit a forecast that differs from their true judgement.

Within this framework, we first find that the FSS can be improved by over-predicting the mean of the neighbourhood fractions (i.e.~by over-predicting the intensity or coverage), which is not true for the BDnSS. For this reason, we advocate only using the FSS such that biases in the mean frequency are removed, such as by using percentile thresholds. This has previously been suggested in e.g.~\cite{skok_estimating_2018}, although this was motivated by making the FSS more accurate for measuring forecast displacement, rather than forecast skill. For forecasters interested in errors in the mean frequency, or interested in truly binary events, then we conclude that the BDnSS is more appropriate as a verification tool. 

The response of these scores to multiplicative errors in the standard deviation of the neighbourhood fractions is then explored, to assess to what extent each score mitigates the double penalty problem. Both the FSS and BDnSS are susceptible to the double penalty problem to some extent, in that they both assign higher scores to more uniform forecasts (i.e.~forecasts that under-predict the standard deviation of the neighbourhood fractions). However, we show that the extent to which this occurs is different for the FSS and BDnSS, and show that this is less of a problem for the FSS than the BDnSS, particularly for smaller neighbourhood sizes and where correlations between forecast and observations are positive but not high. Therefore, whilst the FSS is not suitable for assessing errors in the mean frequency, when errors in the mean are removed it can be more effective than the BDnSS at mitigating the double penalty problem, particularly for small neighbourhood sizes and low positive correlations between forecast and observations. 

Whilst we have focused on neighbourhood scores that are established in the weather forecasting literature, there are many other ways to construct a neighbourhood score, particularly once we have expressed the forecast and observation in terms of summary statistics. We finish by discussing several scores that have appeared in other contexts, the Structural Similarity Index \citep{wang_image_2004}, Kling-Gupta Efficiency \citep{gupta_typical_2011}, and the Symmetric Bounded Efficiency \citep{casati_scale_2023}. These scores are all structured such that, under the framework used here, they do not appear to be susceptible to the double penalty problem at all. The strengths, weaknesses, and interpretations of these scores is an interesting future avenue of research that we leave for future work.

The paper is structured as follows: In Section \ref{sec:nbd_defs} we define notation, in Section \ref{sec:double_penalty} we provide an overview of the double penalty problem in relation to this work, in Section \ref{sec:scores_definition} we define the neighbourhood scores, and in Section \ref{sec:results} we show the results of different forecast errors on the neighbourhood scores.

\section{Definitions}
\label{sec:nbd_defs}
In this section we define notation and terminology for the remainder of the paper, and outline how neighbourhood quantities are calculated.

Consider a forecast, observations and a reference forecast, all provided on a regular equidistant rectangular grid. The neighbourhood fields are constructed by first converting the forecast and observed fields into binary data, by applying a suitable threshold, then averaged over a particular neighbourhood, parameterised by the neighbourhood length $n$ (see e.g.~\citet{roberts_scale-selective_2008}). We define the resulting neighbourhood fractions of the forecast, observations, and reference forecast at spatial location $(i,j)$ and time step $t$ as $f(n)_{ijt}, x(n)_{ijt}$, and $c(n)_{ijt}$ respectively.

We follow the definitions detailed in \cite{antonio_derive_2025} for the summary statistics of the neighbourhood fractions. 
The (sample) means of the neighbourhood fractions for a neighbourhood size $n$ are denoted $\langle f(n) \rangle$, $\langle x(n) \rangle$ and $\langle c(n) \rangle$ for the forecast, observations, and reference forecast respectively. The (sample) standard deviations of the neighbourhood fractions are denoted as $s_{f,n}^2$ $s_{x,n}^2$ and $s_{c,n}^2$. The correlation of the neighbourhood fractions $r_n$ between forecasts and observations is defined as the Pearson correlation between forecast and observation fractions.


Since the domain is finite, a scheme must be chosen to define how neighbourhoods aggregate points near the edges, such as using padding or an enlarged domain \citep{skok_analysis_2016}. The particular choice of edge treatment can have significant effects on the results, particularly at large neighbourhood sizes \citep{skok_analysis_2016}. In this work we assume reflective padding is used in the calculation of the fractions, so that the mean of the neighbourhood fractions is equal to the event frequency at the grid scale, i.e.~$\langle x(n) \rangle = \langle x(0) \rangle$ and $\langle f(n) \rangle = \langle f(0) \rangle$ \citep{antonio_derive_2025}. This enables a clearer interpretation of how values such as the coefficient of variation vary with neighbourhood size, using the results in \cite{antonio_derive_2025}. Other types of edge treatment that use values drawn from the same data distribution will also approximately hold this relation (i.e.~$\langle x(n) \rangle \approx \langle x(0) \rangle$), such as padding with a larger domain and periodic boundary conditions (under the assumption the phenomenon is not too heterogeneous over the domain). We would also expect the results to hold for zero padding, based on empirical observations of the coefficient of variation with neighbourhood size, but do not have a mathematical derivation of this case to assert this rigourously.

\section{Analysis Methodology}
\label{sec:double_penalty}
In this section, we outline the approach used here to investigate how the two scores reflect different types of forecast error. Our approach is based on first expressing the scores in terms of summary statistics, and then transforming these equations to be in terms of relative quantities.

We illustrate our approach with an example. Consider a forecast $f_{i}$, observations $x_{i}$, and a reference forecast $c_{i}$, assumed for simplicity to occur in  a one-dimensional domain. A skill score based on the mean square error of the forecast is the the Nash-Sutcliffe Efficiency, defined as \citep{wilks_forecast_2019}:
\begin{align}
    \text{NSE}(f,x) = 1 - \frac{\sum_i (f_i - x_i)^2}{\sum_i (c_i - x_i)^2}
\end{align}

We begin by expressing the numerator and denominator in terms of the sample mean and standard deviation as performed in \citet{antonio_derive_2025}. We denote the baseline NSE score in the denominator as $B$. We obtain that:
\begin{align}
\label{eq:basic_ss}
    \text{NSE}(f,x) = 1 - \frac{(\langle f \rangle - \langle x \rangle)^2 + s_f^2 + s_x^2 -2rs_fs_x}{B}
\end{align}
For a given value of the error in the mean $(\langle f \rangle - \langle x \rangle)^2$, we can find the stationary points the NSE with respect to $s_{x}$ by solving $\partial (\text{NSE})/\partial s_f =0$. This yields a stationary point at $s_x=r$, and since $\partial^2 (\text{NSE})/\partial s_f^2 = -2/B <0$, this is a maximum. Denoting the maximum $s_{f,\max}$, and constraining $s_{f, max} \ge 0$ since the standard deviation cannot be negative, this can then be written:
\begin{align}\label{eq:sfmax1}
s_{f,\max} := \argmax_{s_f\ge 0}(NSE) = \max(0,rs_x)
\end{align}
In other words, for situations where the forecast and observations are not perfectly correlated, the optimal forecast according to the NSE is one that has $s_f < s_x$, and so is more uniform (closer to the mean) than the observations. This is an example of the `double penalty' problem: if a forecast correctly predicts rain, but misplaces where the rain will occur, then many scores will doubly penalise the forecast; once for predicting rain in the wrong place, and once for not predicting rain where it was actually observed \citep{wilks_forecast_2019}. A score that suffers from the double penalty problem tends to favour predicting more uniform rainfall over a larger area than a smaller patch of rainfall with a more realistic variability; equivalently, the score can be artificially increased when the standard deviation is decreased, as seen in eq.~\eqref{eq:sfmax1}. This framing of the double penalty problem is similar to the definition given in \cite{subich_fixing_2025}.

For our analysis, we are interested in investigating how scores will behave as a function of changes in the forecast relative to the observations, rather than absolute changes in the forecast. A convenient method of obtaining such an expression for the scores we consider here (which only contain quadratic terms) is to transform the score to an equivalent form by dividing both numerator and denominator by $\langle x \rangle^2$. For the skill score in eq.~\eqref{eq:basic_ss}, this gives:
\begin{align}
\label{eq:basic_ss_relative}
    \text{NSE}(f,x) &= 1 - \frac{ \left(\frac{\langle f \rangle}{\langle x \rangle} - 1 \right)^2 + \frac{s_f^2}{\langle x \rangle^2} + \frac{s_x^2}{\langle x \rangle^2} -2r \frac{s_fs_x}{\langle x \rangle^2}}{\frac{B}{\langle x \rangle^2}} := 1 - \frac{ \left(R_\mu - 1 \right)^2 + C^2( R_\sigma^2 + 1 - 2r R_\sigma )}{\tilde{B}}
\end{align}
where $R_\mu := \langle f \rangle / \langle x \rangle, R_\sigma := s_f/s_x$ are ratios of the means and standard deviations respectively, $C := s_x/ \langle x \rangle$ is the coefficient of variation, and $\tilde{B} := B/\langle x \rangle^2$. $R_\mu, R_\sigma$ quantify the multiplicative error in the mean and standard deviation, and so for a perfect forecast $R_\mu=R_\sigma=1$. In this notation, for a given value of the multiplicative error in the mean $R_\mu$, then the skill score is maximised at $R_{\sigma,\max}$, where:
\begin{align}
R_{\sigma,\max} := \argmax_{R_\sigma \ge 0}(NSE) = \max(0,r)
\end{align}

For our analysis, we analyse the scores with the mean, standard deviation, and correlation of the neighbourhood fractions varied independently of each other. A natural question that arises is to what extent this independent variation is realistic. For correlation, this can be seen to be possible from the definition of (Pearson) correlation itself, where covariances are expressed relative to the mean and standard deviations (although this does not rule out that the correlation may covary with the mean, as discussed below). The standard deviation can be expressed as a product of neighbourhood standard deviation at the grid scale, and a term quantifying correlations between neighbours \citep{antonio_derive_2025}. Varying the mean of the neighbourhood fractions without varying the standard deviation corresponds to a situation where errors in the frequency at the grid scale are compensated for in the forecast by adjusting the strength of correlations between grid cells. This may correspond to forecasts with less structural realism, but not in any way that can be detected by the FSS or BDnSS. Varying the standard deviation of the neighbourhood fractions without varying the mean corresponds to a situation where only the correlations between grid cells are modified, such that there is reduced spatial realism that is detectable by the FSS or BDnSS (e.g.~less coherent aggregation of rainfall objects). We accept however, that there are many situations where the mean, standard deviation and correlation will covary, but we are unaware of any robust theoretical relationships that could be used to constrain the values for this study.

\section{Definition of scores}
\label{sec:scores_definition}
In this section the two scores under consideration are defined. To reduce the number of parameters and to arrive at results that are more widely applicable, we will express each score in terms of quantities relative to the observed event frequency at the grid scale, $\langle x(0) \rangle$. We define the multiplicative error in the frequency $R_\mu := \langle f(0) \rangle/\langle x(0) \rangle$, the multiplicative error in the standard deviation of the neighbourhood fractions $R_\sigma := s_{f,n}/s_{x,n}$, and the coefficient of variation $C=s_{x,n}/\langle x(0) \rangle$. For a forecast with no errors detectable by the FSS and BDnSS, $R_\mu = R_\sigma = r_n=1$. Examples of errors that may still be present when $R_\mu = R_\sigma = r_n=1$ are where the true neighbourhood fractions do not follow a Gaussian distribution (e.g.~by having more weight in the tails, or having a skewed distribution), such that the standard deviation and mean do not fully constrain the true distribution. 

From the derived form of the neighbourhood standard deviation in \cite{antonio_derive_2025}, neighbourhood standard deviation is inversely proportional to $N(N+1)$, where $N$ is the neighbourhood width, multiplied by a summation over pairwise covariance terms. Since $N(N+1)$ is the number of pairs of sites within the neighbourhood, then for cases where spatial correlations decay with distance (which occurs in most realistic cases) this means that $C$ decreases as the neighbourhood size grows.

\subsection{Fractions Skill Score}
\label{sec:fss}

The Fractions Skill Score can be written as \citep{antonio_derive_2025}:
\begin{align}\label{eq:fss_exp}
\text{FSS}(n) & = \frac{2(\langle x(n) \rangle \langle f(n) \rangle + s_{x,n} s_{f,n} r_n)}{\langle x(n) \rangle^2 + \langle f(n) \rangle^2 + s_{x,n}^2 + s_{f,n}^2}
\end{align}
where possible values of the FSS are between 0 (worst) and 1 (ideal).

By dividing numerator and denominator by $\langle x(n) \rangle^2$ we arrive at an expression in terms of $R_\mu, R_\sigma$ and $C$:
\begin{align}
\label{eq:fss_relative}
    \text{FSS}(R_\mu, R_\sigma, C, r_n) =  \frac{2(\frac{\langle f(n) \rangle}{\langle x(n) \rangle} + \frac{s_{x,n}}{\langle x(n) \rangle} \frac{s_{f,n}}{\langle x(n) \rangle} r_n)}{1 + \frac{\langle f(n) \rangle^2}{\langle x(n) \rangle^2} + \frac{s_{x,n}^2}{\langle x(n) \rangle^2} + \frac{s_{f,n}^2}{\langle x(n) \rangle^2}} =\frac{2(R_\mu + R_\sigma C^2 r_n)}{1 + R_\mu^2 + C^2(1 + R_\sigma^2)} 
\end{align}

\subsection{Brier Divergence Skill Score}

The Brier Divergence Skill Score (BDnSS), recently introduced in \cite{stein_evaluation_2024}, uses a similar neighbourhood scheme as the FSS but with a difference in the reference score used. There are also differences in how the neighbourhoods are interpreted; in the BDnSS the use of neighbourhoods is interpreted as resampling the probability of an event at that point, whereas in the FSS there is typically no probabilistic interpretation (apart from the extension of the FSS in \cite{duc_spatial-temporal_2013} to incorporate ensemble forecasts). We put these interpretational differences aside for now and explore how the mathematical structure of the BDnSS behaves compared to that of the FSS.

The FSS and BDnSS start from a similar definition, but with different constructions for the reference forecast. For the FSS, the reference forecast is constructed from the forecast being evaluated \citep{roberts_scale-selective_2008}, in contrast to how skill scores are typically constructed (see e.g.~\cite{wilks_forecast_2019}). The BDnSS however uses a reference forecast of climatology or similar, and so is more in line with the standard definitions of skill scores.

The BDnSS is defined as:
\begin{align}
    \text{BDnSS}(n) &= 1 -  \frac{\sum_{t=1}^{T}\sum_{i=1}^{N_x} \sum_{j=1}^{N_y} ( f(n)_{ijt}  - x(n)_{ijt})^2}{\sum_{t=1}^{T}\sum_{i=1}^{N_x}\sum_{j=1}^{N_y} ( c(n)_{ijk}  - x(n)_{ijt})^2} 
\end{align}
Performing the same expansion in terms of means and variances as performed for the FSS in the previous subsection:
\begin{align}
\label{eq:bdss_2}
    \text{BDnSS}(n)
    &=1 - \frac{ (\langle f(n) \rangle - \langle x(n) \rangle)^2 + s_{f,n}^2 + s_{x,n}^2 - 2r_n s_{f,n}s_{x,n}}{(\langle c(n) \rangle - \langle x(n) \rangle)^2 + s_{c,n}^2 + s_{x,n}^2 - 2s_{c,n}s_{x,n} r_{c,n}}
\end{align}

Dividing eq.~\eqref{eq:bdss_2} by $\langle x(n) \rangle^2$ then produces a form for the BDnSS in terms of relative quantities, that is equivalent to eq.~\eqref{eq:basic_ss_relative}:
\begin{align}
\label{eq:bdnss_relative}
    \text{BDnSS}(R_\mu, R_\sigma, C, r_n) &=1 - \frac{(1 - R_\mu)^2 +  C^2 (1 + R_\sigma^2 - 2r_n R_\sigma)}{B}  
\end{align}
where $B$ is the score for the reference forecast.

\section{Effects of errors on the scores}
\label{sec:results}

It is straightforward to see from eqs.~\eqref{eq:fss_relative} and \eqref{eq:bdnss_relative} that if two forecasts only differ in correlation, the one with higher correlation will have a higher score. We therefore focus on exploring how each score changes with the multiplicative errors $R_\mu$ and $R_\sigma$. This is analysed for each type of error separately, whilst keeping other errors fixed. We desire that a verification score cannot be improved by making the forecast less realistic by increasing errors in some way, similar to the concept of `hedging' for probabilistic forecasts \citep{murphy_hedging_1973}. It is therefore desirable to find a score that ranks a forecast higher than another if it has multiplicative errors $R_\mu,R_\sigma$ closer to 1, when all other factors are held equal. 


\subsection{Effect of errors in the event frequency}
\label{sec:mu_bias}

We start by investigating how the neighbourhood scores outlined in Sec.~\ref{sec:scores_definition} respond to multiplicative errors $R_\mu$ in the neighbourhood frequency. From eq.~\eqref{eq:bdnss_relative} is is straightforward to see that the BDnSS is always maximised when $R_\mu=1$, when all other terms are kept fixed, and so cannot be artificially improved by neighbourhood mean. Therefore we focus on the behaviour of the FSS with changing $R_\mu$, for given values of $C, r_n$ and $R_\sigma$.

\begin{figure}[h]
      \centering
      \includegraphics[width=0.8\textwidth]{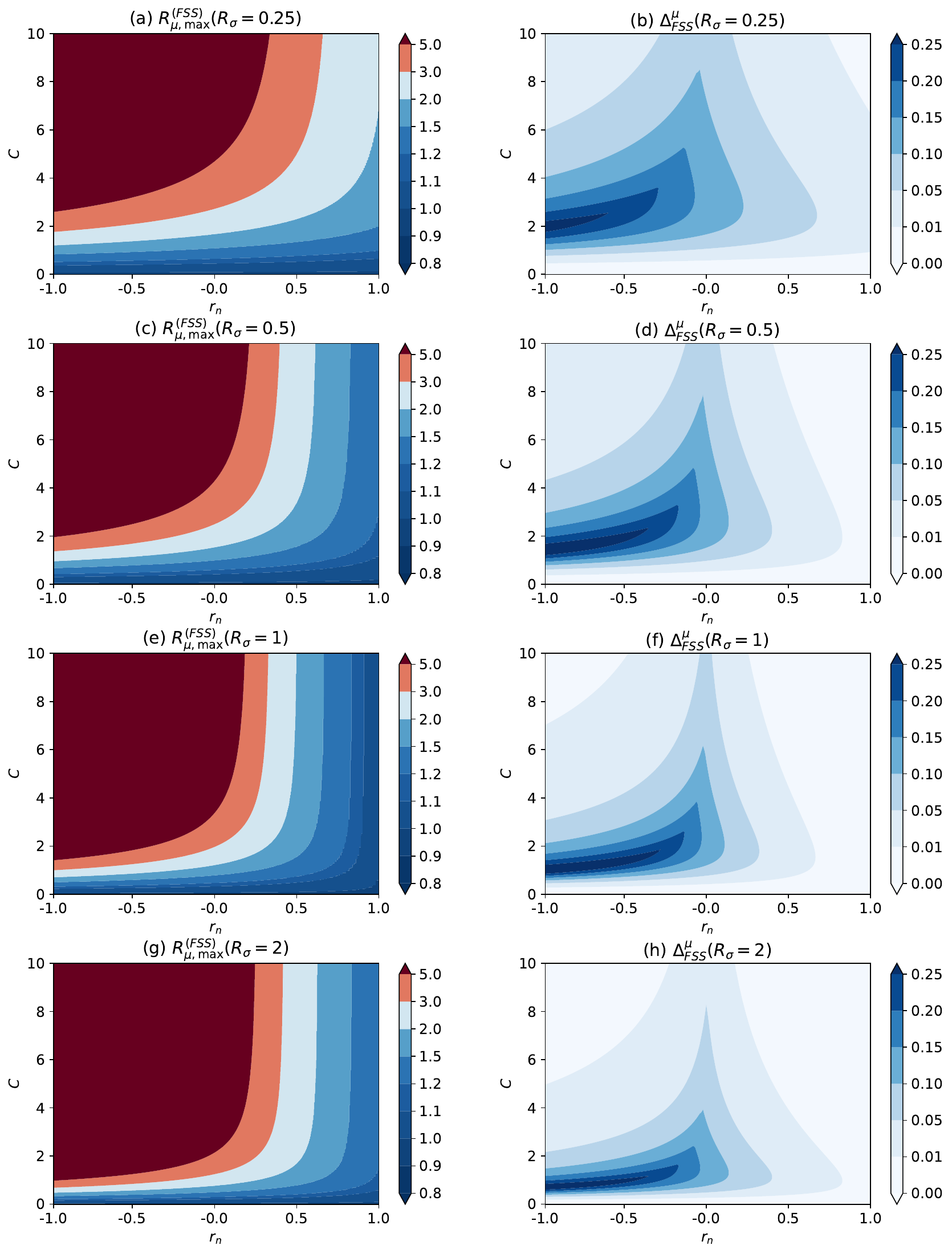}  
     \caption{Analysis of the frequency error that maximises the FSS, for a range of values of the neighbourhood correlation $r_n$, coefficient of variation $C$, and $R_\sigma$. (a), (c), and (e): value of multiplicative error that maximises the FSS, where an ideal verification score has $R_{\mu, max}=1$. (b), (d), and (f): difference between the maximum value of FSS, and the FSS with no error in the frequency (i.e.~$R_{\mu}=1$). $R_\sigma$ increases from top to bottom.}
     \label{fig:bmu_max_comparison}
\end{figure}

By partial differentiation of eq.~\eqref{eq:fss_relative} with respect to $R_\mu$, we identify that the value of $R_\mu$ that maximises the FSS, $R_\mu^{\text{max}}$, is (Appendix A):
\begin{align}
\label{eq:fss_deriv_mu_main}
R_\mu^{\text{max}} = \left[ 1 + C^2(1+R_\sigma^2) + R_\sigma^2 C^4 r_n^2 \right]                                                       ^{1/2} - R_\sigma C^2 r_n.
\end{align}
When $r_n\approx1$ and $R_\sigma\approx1$ (i.e.~for a forecast closely aligned with observations) or for $C\approx0$ (i.e.~large neighbourhood sizes or data with low variance relative to the mean), then $R_\mu^{max}\approx1$. However in general $R_\mu^{max}\neq1$. A numerical evaluation of $R_\mu^{max}$ is shown in Fig.~\ref{fig:bmu_max_comparison} (panels a, c, e and g) for different values of $R_\sigma, C$ and $r_n$, from which it is clear that $R_\mu^{max} \geq 1$, i.e.~the FSS favours forecasts that over-predict the event frequency. From panels a, e, and g we can see that, if $R_\sigma\neq1$, then even with $r_n=1$,  $R_\mu^{\text{max}}>1$, and over-prediction is rewarded.

To understand how this might affect the comparison of two forecasts, we must also quantify the magnitude of the difference in FSS between a forecast with no frequency error and a forecast that maximises the FSS. If the magnitude is small, then the effect may be insignificant compared to variations due to sampling variability. We define $\Delta_{FSS}^{\mu}$ as the difference in score between the FSS with $R_\mu=R_\mu^{max}$ and the FSS with $R_\mu=1$ (no frequency error):
\begin{align}
    \label{eq:fss_delta_mu}
    \Delta_{FSS}^{\mu} := \text{FSS}(R_\mu=R_\mu^{max}, R_\sigma, C, r_n) - \max(0,\text{FSS}(R_\mu=1, R_\sigma, C, r_n))
\end{align}
The second term in eq.~\eqref{eq:fss_delta_mu} uses the $\max(0, \cdot)$ operator, because in cases where $r_n<0$ and $C > 1$ we observe that the FSS with $R_\mu=1$ can become negative, corresponding to an unrealistic regime; this points to a limitation in our approach, in that the summary statistics are not constrained to always produce valid FSS values. Because the FSS will increase from $R_\mu=1$ to $R_\mu^{max}$, and we observe numerically that $\text{FSS}(R_\mu^{max})>0$ for the range of parameters considered here, replacing these negative values with zero corresponds to finding the value of $R_\mu$ that is closest to 1 without being within the unrealistic regime.

The right hand column of Fig.~\ref{fig:bmu_max_comparison} shows $\Delta_{FSS}^{\mu}$ for a range of values of $C$,  $r_n$ and $R_\sigma$. From this it is clear that differences in the scores of more than 0.05 (corresponding to errors of $\ge 5\%$ since $0\le FSS \le 1$) appear below $r_n=0.5$, and differences are particularly high when $r_n < 0$. The value of $C$ that produces the largest differences depends on $R_\sigma$, peaking at $C\approx1$ for $R_\sigma \ge 1$ and $C=2$ for $R_\sigma=2$. In general, this means that the difference is maximised for intermediate neighbourhood sizes.

This analysis demonstrates that, for two forecasts that only differ in the frequency error $R_\mu$, the FSS(n) can assign a higher score to the forecast that has higher error, except in situations where the forecast is perfectly correlated with observations, or when the neighbourhood size is very large. The magnitude of this effect is particularly high for low correlation values, and intermediate neighbourhood sizes.

Therefore, to ensure the FSS cannot be artificially improved by introducing errors in the neighbourhood frequency, this provides robust motivation for calculating the FSS using percentile thresholds (i.e.~defining events as those that exceed a percentile value rather than an absolute value) to remove the neighbourhood frequency bias, in effect performing quantile mapping on the forecast. This was also recommended in \cite{skok_estimating_2018}, although the motivation was to ensure that the FSS could be used as a method to measure forecast displacement accurately. In addition, the choice of padding used is important; as discussed in \cite{antonio_derive_2025}, neighbourhood frequency biases are not removed using percentile thresholds with zero padding, however they are when using percentile thresholds with e.g.~reflective padding. In some cases percentile thresholds may not be desirable or cannot be used (such as when forecasting truly binary events), and therefore in these cases these results motivate the use of the BDnSS in place of the FSS.

\subsection{Effects of errors in the standard deviation of the neighbourhood fractions}
\begin{figure}[h]
      \centering
      \includegraphics[width=0.8\textwidth]{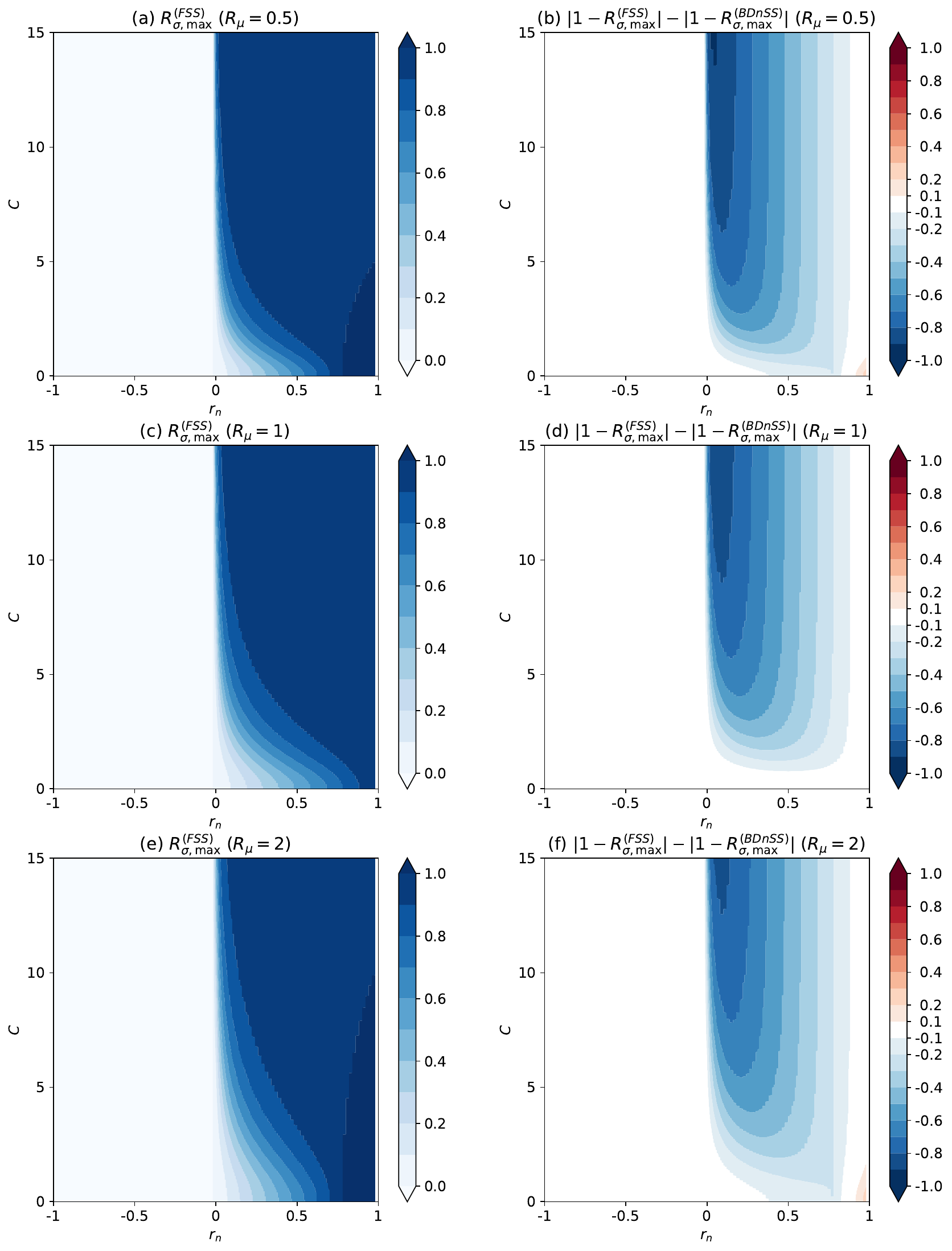}  
     \caption{Comparison of values of $R_\sigma$ that maximise the FSS and BDnSS scores, for different values of the coefficient of variation $C$ and neighbourhood correlation $r_n$ (a), (c) and (e): the value of $R_\sigma$ that maximises the FSS, for different values of $R_\mu$; (b), (d) and (f): the difference in $|1-R_{\sigma, \max}^{ (\text{FSS})}|$ and $|1-R_{\sigma, \max}^{ (\text{BDnSS})}|$: negative values indicate where the FSS is less susceptible to the double penalty problem than the BDnSS.}
     \label{fig:bsigma_max_comparison}
\end{figure}

In this section we investigate how the verification scores respond to multiplicative errors in the standard deviation of the neighbourhood fractions.
For the BDnSS, since it has the same form as the MSE skill score presented in Sec.~\ref{sec:double_penalty}, we see that BDnSS is maximised  at $R_{\sigma, \max}^{ (\text{BDnSS})}$, where:
\begin{align}
    R_{\sigma, \max}^{ (\text{BDnSS})} := \argmax_{R_\sigma \geq 0}\left(\text{BDnSS}\right) = \max(0, r_n)
\end{align}
We therefore see that the BDnSS assigns higher scores to forecasts with lower variance, when the correlation is less than 1, as expected since it is a score constructed using squared errors. From eq.~\eqref{eq:bdnss_relative} it is clear that, as $C$ decreases (corresponding to an increase in neighbourhood size), the effect that $R_\sigma$ has on the overall BDnSS score decreases. 

By calculating the partial derivative of eq.~\eqref{eq:fss_relative} with respect to $R_\sigma$ and setting this to zero with $R_\mu=1$, we find that the FSS is maximised by $R_{\sigma, \max}^{ (\text{FSS})}$ where (see Appendix A):
\begin{align}
\label{eq:fss_bsigma_main}
    R_{\sigma, \max}^{ (\text{FSS})} =\argmax_{R_\sigma \geq 0}\left(\text{FSS}\right) =\max\left(0,\frac{1}{C^2r_n} \left[ R_\mu^2 + C^2 r_n^2 (1 + R_\mu^2 + C^2)\right]^{1/2} - \frac{R_\mu}{C^2r_n} \right)
\end{align}
Particularly interesting limits of this are for small values of $C$ (which corresponds to large neighbourhood sizes, which reduce the variance relative to the mean) and large $C$ (which corresponds to small neighbourhood sizes, and/or where there is a large variance relative to the mean). In the limit of large $C$:
\begin{align}
\lim_{C \to \infty} \left(R_{\sigma, \max}^{ (\text{FSS})}\right) = 1
\end{align}
This means that, in the limit of small neighbourhood sizes, the FSS is maximised at $R_\sigma=1$, unlike the BDnSS. In the limit of small $C$ when $R_\mu = 1$:
\begin{align}
\lim_{C \to 0, R_\mu=1} \left(R_{\sigma, \max}^{ (\text{FSS})}\right) = \max\left(0,r_n\right)
\end{align}
This means that, in the limit of large neighbourhood sizes ($C\to 0$), the properties of the FSS and BDnSS converge when $R_\mu=1$.

Numerical values of $R_{\sigma, \max}^{ (\text{FSS})}$ are plotted in Fig.~\ref{fig:bsigma_max_comparison} in order to assess how it behaves for different values of $R_\mu$, $r_n$, and $C$. For the majority of values of $R_\mu$, $r_n$, and $C$, $R_{\sigma, \max}^{ (\text{FSS})}$ lies between 0 and 1, as is also the case for $R_{\sigma, \max}^{ (\text{BDnSS})}$; both scores can therefore be artificially increased by under-predicting the neighbourhood variance, i.e.~by making forecasts more uniform and closer to the mean. As discussed in Sec.~\ref{sec:double_penalty}, an interpretation of this is that both scores suffer from the double penalty problem. An exception to this generalisation can be seen for the FSS in Fig.~\ref{fig:bsigma_max_comparison} (a) and (e), when $R_\mu\neq1$ and $r_n\approx 1$; in this regime, the FSS favours forecasts that \emph{over-predict} the standard deviation of the neighbourhood fractions.

Values of $|1-R_{\sigma, \max}^{ (\text{FSS})}| - |1-R_{\sigma, \max}^{ (\text{BDnSS})}|$ are plotted in Fig.~\ref{fig:bsigma_max_comparison} (b): areas where this quantity is negative indicate where the FSS is less susceptible to artificial score improvements by reducing the neighbourhood variance of the forecast. For all values of $R_\mu,r_n$ and $C$, we see mostly that $|1-R_{\sigma, \max}^{ (\text{FSS})}| - |1-R_{\sigma, \max}^{ (\text{BDnSS})}| \leq 0$, with more pronounced differences particularly for lower correlations and higher $C$ (small neighbourhood sizes and / or high variance relative to the mean). This shows that, in general, the FSS favours forecasts that are less biased in the standard deviation of the neighbourhood fractions. Under the interpretation of the double penalty problem given in Sec.~\ref{sec:double_penalty}, the FSS appears generally less susceptible to the double penalty problem than the BDnSS, in that it is generally maximised for smaller biases in standard deviation. The difference is particularly high when the neighbourhood correlation is low, neighbourhood sizes are small, or variance relative to the mean is high.

As with the analysis in the previous subsection, we can check the magnitude of the difference between the maximised score and the score with no standard deviation error. In order to estimate this, we assume a particular baseline forecast for the BDnSS. We assume a reference score that is uncorrelated with the observations ($r_n=0$), with $R_{\mu}=1$ and $R_{\sigma}=1$ (i.e. otherwise well-calibrated). This is a baseline that is harder to beat than a pointwise random forecast, which would have $R_{\sigma} <1$ due to lack of spatial correlations, but may produce similar results to using climatology as the reference forecast, which would have $R_\mu\approx1$, $R_\sigma <1$ (more uniform) and a small but non-zero $r_n$. With this baseline, the denominator in eq.~\eqref{eq:bdnss_relative} becomes $2C^2$. As in the previous subsection, we define $\Delta_{\text{FSS}}^{\sigma}, \Delta_{\text{BDnSS}}^{\sigma}$ as:
\begin{align}
\label{eq:fss_delta_sigma}
    \Delta_{\text{FSS}}^{\sigma} &:= \text{FSS}(R_\mu, R_\sigma=R_{\sigma,\max}^{(\text{FSS})}, C, r_n) - \text{FSS}(R_\mu, R_\sigma=1, C, r_n) \\
    \Delta_{\text{BDnSS}}^{\sigma} &:= \text{BDnSS}(R_\mu, R_\sigma=R_{\sigma,\max}^{(\text{BDnSS})}, C, r_n) - \text{BDnSS}(R_\mu, R_\sigma=1, C, r_n)\label{eq:bdnss_delta_sigma}
\end{align}

\begin{figure}[h]
      \centering
      \includegraphics[width=\textwidth]{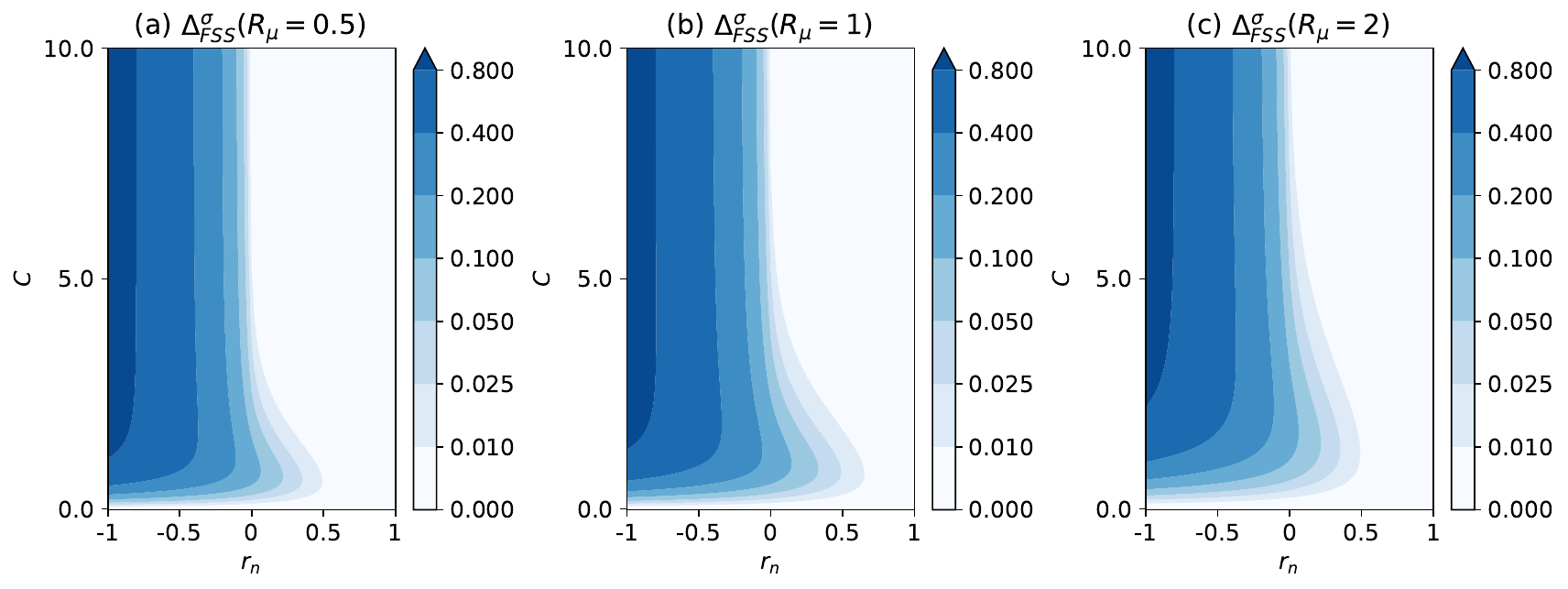}  
     \caption{Analysis of difference between maximised scores, and scores with no error in the standard deviation of the neighbourhood fractions (i.e. $R_{\sigma}=1$) for the FSS, for different values of $C$ and $r_n$. Note the nonlinear colorscale.}
     \label{fig:bsigma_max_effect_fss}
\end{figure}

$\Delta_{\text{FSS}}^{\sigma}$ is plotted in Fig.~\ref{fig:bsigma_max_effect_fss}. This is generally small for forecasts with positive neighbourhood correlation with observations ($r_n>0$), but grows larger when $r_n<0.5$ and $C>1$. The effects are particularly pronounced for negative neighbourhood correlations, and increasing $R_\mu$ generally has the effect of increasing the differences at low correlations and intermediate values of $C$.  

\begin{figure}[h]
      \centering
      \includegraphics[width=0.5\textwidth]{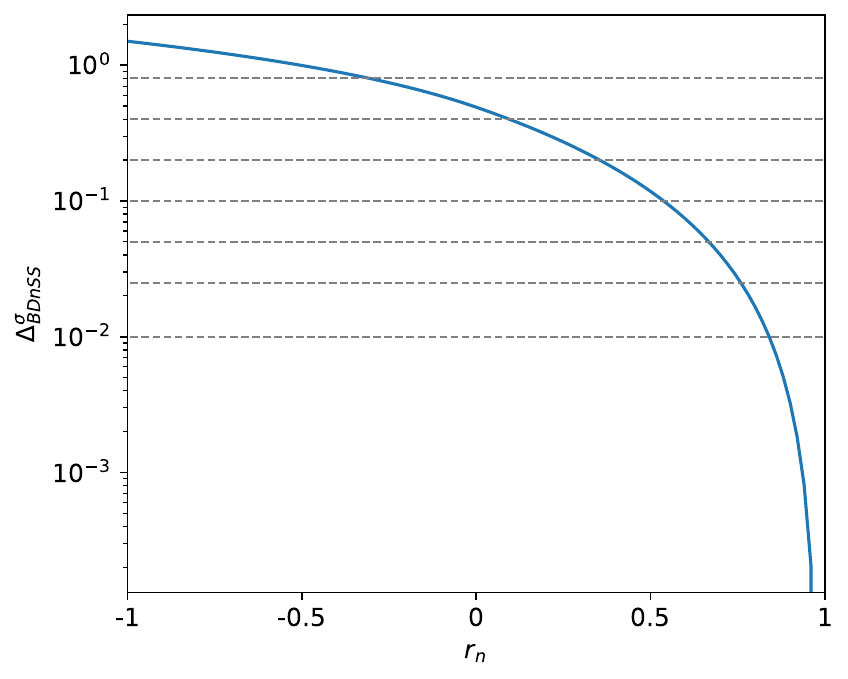}  
     \caption{As for Fig.~\ref{fig:bsigma_max_effect_fss} but for the BDnSS. Dotted lines show the colorbar levels in Fig.~\ref{fig:bsigma_max_effect_fss}.}
     \label{fig:bsigma_max_effect_bdnss}
\end{figure}

Fig.~\ref{fig:bsigma_max_effect_bdnss} shows $\Delta_{\text{BDnSS}}^{\sigma}$, where only the dependence on $r_n$ is shown because eq.~\eqref{eq:bdnss_delta_sigma} does not depend on $R_\mu$ or $C$ for the reference score we have chosen. For $r_n =0.5$, $\Delta_{\text{BDnSS}}^{\sigma} \approx 0.1$, around 10 times larger than the differences of $\lesssim 0.01$ seen for the FSS in Fig.~\ref{fig:bsigma_max_effect_fss}. $\Delta_{\text{BDnSS}}^{\sigma}$ takes values of around 0.1-0.5 for $r_n>0$; since the BDnSS lies between 0 and 1 for $r_n>0$, this suggests that this effect would be significant compared to variations due to sampling uncertainty.

At first this may appear at odds with the results demonstrated in \cite{stein_evaluation_2024}, which showed that the BDnSS is maximised on synthetic data when biases in standard deviation at the grid scale are minimised. However, the synthetic data used in their example was such that forecast and observations share a common background, and differences are only in the local noise applied to this background. Therefore the neighbourhood correlation in their example is very close to 1, and so the score is maximised when the multiplicative error in the standard deviation of the neighbourhood fractions $R_\sigma$ is 1, in agreement with this analysis.

\section{Discussion and Conclusions}
\label{sec:conc}

In this work we have explored how errors in the mean and standard deviation of the neighbourhood fractions are reflected in two neighbourhood scores; the Fractions Skill Score (FSS), which is perhaps the most commonly used neighbourhood score, and the newly proposed Brier Divergence Skill Score (BDnSS), which differs in using a reference forecast that is independent of the forecast, as is more conventional for skill scores. By expressing each score in terms of multiplicative errors in the mean and standard deviation of the neighbourhood fractions, we explore how each score will rank forecasts with different errors, and to what extent the scores can be artificially improved by increasing one of these errors. 

Using this approach, we first demonstrate that, unlike the BDnSS, the FSS may assign a higher score to forecasts that over-predict the event frequency. This provides more rigourous motivation to only use this score with percentile thresholds, as previously recommended in \cite{skok_estimating_2018}. We then analyse how each score reflects errors in the forecasted standard deviation of the neighbourhood fractions. We demonstrate that both the FSS and BDnSS can assign higher scores to forecasts that under-predict standard deviation of the neighbourhood fractions, which is partly a simple reflection of the well-known tendency for scores based on mean-squared error to score more uniform predictions higher when forecast and observations are not perfectly correlated. However, we find the extent to which the FSS favours more uniform predictions is less pronounced than for the BDnSS, particularly for situations where the correlation between forecast and observations is lower and positive, or where the coefficient of variation is high (corresponding to small neighbourhoods, or highly variable data). We conclude that, for particular situations, the FSS can be less susceptible to the double penalty problem than the BDnSS.

A limitation of this approach is that it does not account for realistic ways that neighbourhood mean, standard deviation, and correlation can vary together, and can produce combinations of summary statistics that are unlikely to occur, or that can be unphysical. Constraining the search space by using empirical or theoretical relationships between mean, standard deviation and correlation is therefore a worthwhile avenue for improving this analysis.

The differing behaviours of the FSS and BDnSS are due to the particular ways in which the mean, standard deviation, and correlation of the neighbourhood fractions are combined to construct a summary value, and the different ways in which these scores use reference forecasts. There are many other functional forms we may consider which may have better properties than either the FSS or BDnSS. Particularly interesting scores are the Structural Similarity Index (SSIM, \citet{wang_image_2004}), Kling-Gupta Efficiency (KGE, \citet{gupta_typical_2011}), and Symmetric Bounded Efficiency (SBE, \citet{casati_scale_2023}). Using our notation, these scores are:
\begin{align}
\text{SSIM} &= \left[ \frac{2 \langle x \rangle \langle f \rangle + \beta_1}{(\langle x \rangle^2 + \langle f \rangle^2) + \beta_1}\right]^{\alpha_1} \left[ \frac{2s_x s_f + \beta_2}{(s_x^2 + s_f^2) + \beta_2}\right]^{\alpha_2} \left[\frac{r s_x s_f + \beta_3}{s_x s_f + \beta_3} \right]^{\alpha_3} \\
\text{KGE} &= 1 - \sqrt{(r-1)^2 + \left( \frac{\langle x \rangle}{\langle f \rangle }-1 \right)^2 + \left( \frac{s_x}{s_y} -1 \right)^2} \\
\text{SBE} &= 1 - \sqrt{(r-1)^2 + \left( \frac{\langle x \rangle - \langle f \rangle }{\langle x \rangle + \langle f \rangle} \right)^2 + \left( \frac{s_x - s_f }{s_x + s_f} \right)^2}
\end{align}
where $\beta_1, \beta_2, \beta_3$ are small constants to ensure that the SSIM is not undefined when values are close to 0. Because these scores all involve separate terms in the mean, standard deviation and correlation, it is straightforward to see that these scores all satisfy the desired behaviour, that a score is maximised at the point with minimal error, all other factors being equal (a subtlety is that this is only true for the SSIM when $r>0$, however it is easy to modify the SSIM to use $\frac{1}{2}(1+r)$ instead to extend this property to all correlation values). However, these scores may lack interpretability compared to scores like the FSS and BDnSS, and are relatively unexplored in the context of weather verification. The free parameters in the SSIM also raise the question of how to constrain these parameters in a principled, interpretable way. An exploration of the properties of these scores therefore presents a promising avenue of future research. 




\clearpage
\acknowledgments
This publication is part of the EERIE project funded by the European Union (Grant Agreement No 101081383). Views and opinions expressed are however those of the author(s) only and do not necessarily reflect those of the  European Union or the European Climate Infrastructure and Environment Executive Agency (CINEA). Neither the European Union nor the granting authority can be held responsible for them. This work was funded by UK Research and Innovation (UKRI) under the UK government’s Horizon Europe funding guarantee (grant number 10049639). 

%
%
\datastatement

The Python code and data used to create the plots in this work can be found at \url{https://github.com/bobbyantonio/fractions_skill_score}.

\appendix[A]
\appendixtitle{Partial derivatives of the FSS}
\label{app:fss_bmu}
In this appendix we provide the derivation of the partial differentiation of the FSS, as expressed in eq.~\eqref{eq:fss_relative}, with respect to $R_\mu$ and $R_\sigma$. The FSS equation is:
\begin{align}
    \text{FSS}(n) = \frac{2(R_\mu + R_\sigma C^2 r_n)}{1 + R_\mu^2 + C^2(1 + R_\sigma^2)} 
\end{align}

We start by calculating the derivative with respect to $R_\mu$. To enable a simpler derivation, we first group any terms that do not explicitly depend on $R_\mu$ into variables $\alpha:=R_\sigma C^2 r_n, \beta:=1 + C^2(1 + R_\sigma^2)$:
\begin{align}
    \text{FSS}(n) = \frac{2(R_\mu + \alpha)}{R_\mu^2 + \beta} 
\end{align}
Performing the partial differentiation with respect to $R_\mu$, and setting this to zero, we have:
\begin{align}
    \frac{\partial \text{FSS}(n)}{\partial R_\mu} = \frac{2}{R_\mu^2 + \beta} - \frac{4R_\mu(R_\mu + \alpha)}{(R_\mu^2 + \beta)^2} = 0
\end{align}
We assume that $R_\mu$ is finite so that the denominator cannot bring the derivative close to zero. Therefore we multiply through by $(R_\mu^2 + \beta)^2$ and simplify to get:
\begin{align}
    2R_\mu^2 + 4\alpha R_\mu - 2\beta = 0
\end{align}
Solving this quadratic equation and substituting the full forms of $\alpha, \beta$ back in, we obtain an expression for the unique value of $R_\mu$ that maximises $\text{FSS}(n)$, which we denote $R_\mu^{\text{max}}$:
\begin{align}
\label{eq:bmu_deriv_app}
    R_\mu^{\text{max}} := \argmax_{R_\mu \geq 0}\left(\text{FSS}(n)\right) = \left[ 1 + C^2(1+R_\sigma^2) + R_\sigma^2 C^4 r_n^2 \right]^{1/2} - R_\sigma C^2 r_n
\end{align}
where we have only kept the positive solution to the quadratic equation since $ R_\mu \geq 0$.

Similarly we can perform a partial differentiation of the FSS with respect to $R_\sigma$. As above, we first group any terms that do not explicitly depend on $R_\sigma$ into variables $\alpha:= C^2 r_n, \beta:=1 + C^2 + R_\mu^2$:
\begin{align}
    \text{FSS}(n) = \frac{2(R_\mu + \alpha R_\sigma)}{C^2R_\sigma^2 + \beta} 
\end{align}
Performing the partial differentiation with respect to $R_\sigma$, and setting this to zero, we have:
\begin{align}
    \frac{\partial \text{FSS}(n)}{\partial R_\sigma} = \frac{2\alpha}{C^2R_\sigma^2 + \beta} - \frac{4 C^2 R_\sigma(R_\mu + \alpha R_\sigma)}{(C^2 R_\sigma^2 + \beta)^2} = 0
\end{align}
With the constraint that $R_\sigma$ is finite, we multiply through by $(C^2 R_\sigma^2 + \beta)^2$ and rearrange to obtain the quadratic equation:
\begin{align}
 2\alpha C^2 R_\sigma^2 + 4C^2 R_\mu R_\sigma - 2 \alpha \beta = 0 
\end{align}
solving this equation and substituting for $\alpha,\beta$, we obtain:
\begin{align}
    R_\sigma^{\text{max}} := \argmax_{R_\sigma \geq 0}\left(\text{FSS}(n)\right) = \frac{1}{C^2r_n} \left[ R_\mu^2 + C^2 r_n^2 (1 + R_\mu^2 + C^2)\right]^{1/2} - \frac{R_\mu}{C^2r_n}
\end{align}
Note that the first term in this equation is greater than or equal to the second term. Therefore, when $r_n < 0$, this equation becomes negative. Since $R_\sigma \geq 0$, we must modify the equation to:
\begin{align}
R_\sigma^{\text{max}} = \max\left(0,\frac{1}{C^2r_n} \left[ R_\mu^2 + C^2 r_n^2 (1 + R_\mu^2 + C^2)\right]^{1/2} - \frac{R_\mu}{C^2r_n} \right)
\end{align}


\bibliographystyle{ametsocV6}
\bibliography{references}

\end{document}